\documentclass[showpacs,aps,graphicx,twocolumn]{revtex4}%
\usepackage{graphicx}

\begin{document}

\title{Secure quantum key distribution network with Bell states and local unitary
 operations \footnote{published in \emph{Chinese Physics Letters} \textbf{22} (5),
 1049-1052 (2005).}}
\author{ Chun-Yan Li$^{1,2,3}$, Hong-Yu Zhou$^{1,2,3}$, Yan Wang$^{1,2,3}$, Fu-Guo
Deng$^{1,2,3}$\footnote{ E-mail:
 fgdeng@bnu.edu.cn}}
\address{  $^1$ The Key Laboratory of Beam Technology and Material
Modification of Ministry of Education, Beijing Normal University,
Beijing 100875,
China\\
$^2$ Institute of Low Energy Nuclear Physics, and Department of
Material Science and Engineering, Beijing Normal
University, Beijing 100875,  China\\
$^3$ Beijing Radiation Center, Beijing 100875,  China }
\date{\today }

\begin{abstract}
We propose a theoretical scheme for secure quantum key
distribution network following the ideas in quantum dense coding.
In this scheme, the server of the network provides the service for
preparing and measuring the Bell states, and the users encodes the
states with local unitary operations. For preventing the server
from eavesdropping, we design a decoy when the particle is
transmitted between the users. It has high capacity as one
particle carries two bits of information and its efficiency for
qubits approaches 100\%. Moreover, it is not necessary for the
users to store the quantum states, which makes this scheme more
convenient for application than others.
\end{abstract}
 \maketitle

Quantum key distribution (QKD), the most advanced application of
the principles in quantum mechanics within the field of
information, such as the uncertainty principle and quantum
correlations, provides a secure way for two remote parties, say
Alice and Bob to create a randomly binary string that can be used
as a private key with which they can communicate securely using
Vernam one-time pad crypto-system \cite{vernam}. In the key
transmission with classical line only, a vicious eavesdropper, Eve
can monitor the line freely without leaving a trace. But the
uncertainty principle assures that Eve cannot copy an unknown
quantum state of single particle \cite{nocloning}. In 1984,
Bennett and Brassard \cite{BB84} designed the first theoretical
model for quantum key distribution based on non-cloning theorem
\cite{nocloning}, called BB84 QKD protocol. Quantum correlations
of entangled particles or wave-packets also help people to do key
distribution in an unconditionally secure way. Ekert \cite{Ekert}
proposed another QKD scheme based on the correlation of
Einstein-Podolsky-Rosen (EPR) pair, the maximal entangled state of
two particles in 1991. To date, a great number of works have been
done on QKD both in theoretical aspects
\cite{BBM92,B92,abc,cabello,longliu,guoQKDE,CORE,guiguo,hanguo,BidQKD,delay}
and in experimental implementations \cite{KZHWGTR,QKD122}.

Most of the existing QKD protocols concentrate on point-to-point
key distribution between two remote parties. The practical
application of QKD requires the communication of any-to-any key
distribution on a network, same as the classical communication
network (world web). Unfortunately, there are only a little of
works focused on multi-user quantum key distribution (MUQKD)
\cite{MUQKD1,MUQKD2,MUQKDNature,MUQKDguo,DLMXL} on a passive
optical network. Some \cite{MUQKD1,MUQKD2,MUQKDNature} of them
choose single photons as quantum information carrier (QIC) and
measure them with two sets of measuring bases (MBs), the
rectilinear basis $\sigma_z$ and the diagonal basis $\sigma_x$.
Their total efficiency $\eta_t$ is low. $\eta_t$ is defined as
\cite{cabello,longliu}
\begin{equation}
\eta_t=\frac{b_s}{q_t+b_t}, \label{eff1}
\end{equation}
where $b_s$ is the number of bits in the key, $q_t$ is the number
of qubits used, and $b_t$ is the number of classical bits
exchanged between the parties. For example, the efficiency
$\eta_t$ in Ref. \cite{MUQKD2} is lower than $\frac{1}{16}$ as no
more than $\frac{1}{8}$ QIC can be used as the qubits in the raw
key. Xue et. al. proposed a MUQKD protocol with the combination of
two-particle product states and entangled states following the
ideas in Ref. \cite{abc}, and almost all of the instances can be
used as the raw key and two particles can carry one bit of quantum
information. In the MUQKD scheme \cite{DLMXL}, EPR pairs are used
as QIC and are transmitted with two quantum channels. The four
local unitary operations represent four kinds of coding. It is the
generalization of the Long-Liu point-to-point QKD protocol
\cite{longliu} into the case with many users on a passive optical
network. With quantum storage (quantum memory)
\cite{storage1,storage2,sun,wang}, its efficiency for qubit
$\eta_q\equiv\frac{q_u}{q_t}$ approaches 100\% and its total
efficiency $\eta_t$ approaches 50\% as all the EPR pairs are
useful for the raw key and only two bits of classical information
are exchanged for two qubits, where $q_u$ is the useful qubits.
Certainly, the technique of quantum storage is not fully developed
at present. However it is a vital ingredient for quantum
computation and quantum information, and there has been great
interests in developing it \cite{storage1,storage2,sun,wang}. It
is believed that this technique will be available in the future.
With quantum memory, many new applications can be constructed,
such as quantum computation \cite{book}, quantum secure direct
communication \cite{
twostep,QOTP,yan0,Gaot0,GaoT1,caiqycpl,zhangzj,luh,zhangzj2,wangc}
and quantum secret splitting \cite{HBB99,Peng}.

In this paper, we want to introduce  a MUQKD scheme with EPR pairs
following the ideas in quantum dense coding \cite{densecoding}. In
this scheme, the users on the network need only perform
single-particle measurement and exploit a decoy technique,
replacing some of the particles in the original QIC with those
whose states are unknown for others, to guarantee its security.
The information is encoded on the states with four local unitary
operations. The efficiency and the capacity of this MUQKD scheme
are maximal, same as those in Ref. \cite{DLMXL}. Moreover it does
not require the users to  store the quantum states received and
only one particle in each EPR pair runs through the quantum
channel, which  make this MUQKD scheme more convenient for the
practical application.

An EPR pair is in one of the four Bell states shown as
follows\cite{longliu,CORE,DLMXL}:
\begin{eqnarray}
\left\vert \psi ^{-}\right\rangle_{AB}
=\frac{1}{\sqrt{2}}(\left\vert 0\right\rangle _{A}\left\vert
1\right\rangle _{B}-\left\vert
1\right\rangle _{A}\left\vert 0\right\rangle _{B}), \label{EPR1}\\
\left\vert \psi ^{+}\right\rangle_{AB}
=\frac{1}{\sqrt{2}}(\left\vert 0\right\rangle _{A}\left\vert
1\right\rangle _{B}+\left\vert
1\right\rangle _{A}\left\vert 0\right\rangle _{B}), \label{EPR2}\\
\left\vert \phi ^{-}\right\rangle_{AB}
=\frac{1}{\sqrt{2}}(\left\vert 0\right\rangle _{A}\left\vert
0\right\rangle _{B}-\left\vert
1\right\rangle _{A}\left\vert 1\right\rangle _{B}), \label{EPR3}\\
\left\vert \phi ^{+}\right\rangle_{AB}
=\frac{1}{\sqrt{2}}(\left\vert 0\right\rangle _{A}\left\vert
0\right\rangle _{B}+\left\vert 1\right\rangle _{A}\left\vert
1\right\rangle _{B}). \label{EPR4}
\end{eqnarray}
The four local unitary operations $U_{i}$ ($i=0,1,2,3$) can
transform one of the Bell states into each other.
\begin{eqnarray}
U_{0}=\left\vert 0\right\rangle \left\langle 0\right\vert
+\left\vert 1\right\rangle \left\langle 1\right\vert, \,\,\,\,\,
U_{1}=\left\vert 0\right\rangle \left\langle 1\right\vert
-\left\vert 1\right\rangle \left\langle 0\right\vert, \nonumber\\
U_{2}=\left\vert 1\right\rangle \left\langle 0\right\vert
+\left\vert 0\right\rangle \left\langle 1\right\vert, \,\,\,\,\,\,
U_{3}=\left\vert 0\right\rangle \left\langle 0\right\vert
-\left\vert 1\right\rangle \left\langle 1\right\vert. \label{U}
\end{eqnarray}
For example,
\begin{eqnarray}
I \otimes U _{0}\vert \phi^+\rangle&=&\vert \phi^+\rangle,
\,\,\,\,\,\,\,\, I \otimes U _{1}\vert \phi^+\rangle=-\vert
\psi^-\rangle,\label{L1}\\
I \otimes U _{2}\vert \phi^+\rangle&=&\vert \psi^+\rangle,
\,\,\,\,\,\,\,\, I \otimes U_{3}\vert \phi^+\rangle=\vert
\phi^-\rangle,\label{3}
\end{eqnarray}
where $I=U_0$ is the 2$\times$2 identity matrix.

First, let us compare the quantum dense coding with the Long-Liu
point-to-point QKD scheme. In quantum dense coding
\cite{densecoding}, the QIC is the EPR pairs transmitted in one by
one and one of the two particles in each pair runs forth and back
from  the receiver of information, Carol to the sender Bob. The
other particle is hold in the hand of Carol. The information is
encoded on the state with the four unitary operations $U_i$ chosen
randomly by Bob. After the particle encoded returns to Carol, she
performs the Bell state measurement on the EPR pair and reads out
the information about the operation. In this way, a particle can
carry two bits of information with running forth and back. In the
Long-Liu point-to-point QKD protocol \cite{longliu}, the EPR pairs
are transmitted by using two split channels in a quantum data
block, which is necessary for QSDC \cite{twostep,yan0,QOTP} but
not for QKD as the analysis of the security in QKD is just a
post-processing. The advantage of Long-Liu QKD protocol
\cite{longliu} is that the loss of the qubits  is lower by far
than that in quantum dense coding when there are noise and loss in
the quantum channel as all the QIC are transmitted from the sender
to the receiver once. This advantage will disappear in the case
\cite{DLMXL} with many user on a network.

\begin{figure}[!h]
\begin{center}
\includegraphics[width=6cm,angle=0]{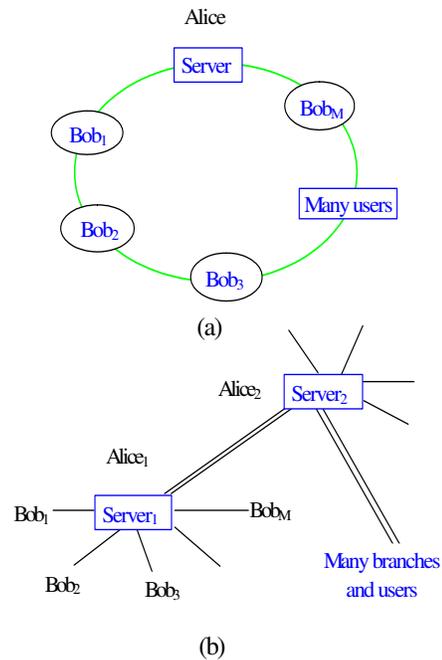} \label{f1}
\caption{ The topological structure of the network, similar to
those in Refs. \cite{MUQKD1,MUQKD2,MUQKDNature,MUQKDguo,DLMXL}:
(a) loop-configuration network; (b) star-configuration network.}
\end{center}
\end{figure}
\begin{figure}[!h]
\begin{center}
\includegraphics[width=7cm,angle=0]{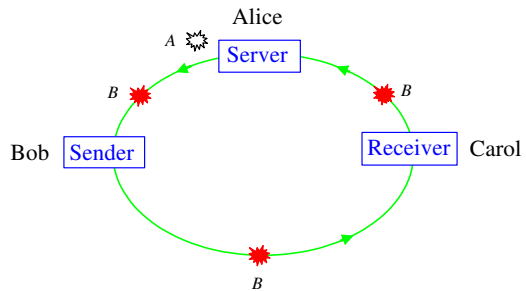} \label{f2}
\caption{ The subsystem of the network in this MUQKD scheme. The
server provides the service for preparing and measuring the Bell
states, Bob and Carol choose randomly the control mode and the
coding mode for the particles received. They sends the particles
to next one when they  choose the coding mode, otherwise they
perform single-particle measurement on the particles received with
one of the two MBs randomly. For preventing the server from
eavesdropping, Bob exploit the decoy technique with a certain
probability, i.e., replacing the original particle with his one
prepared with one of the two MBs. }
\end{center}
\end{figure}
Now we discuss our MUQKD scheme in detail. Although the
topological structure of the network can be loop or star, similar
to those in Refs. \cite{MUQKD1,MUQKD2,MUQKDNature,MUQKDguo,DLMXL}
shown in Fig.1, its subsystem can be simplified to that in Fig.2,
composed of the server (Alice), the sender (Bob) and the receiver
(Carol). Suppose Alice is the server of the sender, Bob. If Carol
is in another branch of the network, her server, say $Alice_{i}$
provides the quantum channel for her to communicate with Bob only
in a given time slot \cite{MUQKDguo,DLMXL}. So this MUQKD scheme
is explicit if we describe clearly the subsystem in Fig.2. For the
integrality of this MUQKD scheme, we describe the steps in detail,
including some same as those in Ref. \cite{DLMXL}.

(S1) All the users on the network agree that the four unitary
operations, $U_0$, $U_1$, $U_2$ and $U_3$ represent the bits 00,
01, 10 and 11, respectively. The server Alice prepares the QIC in
the original state $\vert\phi^+\rangle_{AB}$.

(S2) Alice sends the particle $B$ to Bob and keeps the particle
$A$ in home.

(S3) Bob chooses randomly the control mode or the coding mode,
similar to that in Ref. \cite{bf}. When he chooses the control
mode, he performs the single-particle measurement on particle $B$
by choosing the two MBs, $\sigma_z$ or $\sigma_x$ with the same
probability. He tell Alice his MB for the particle $B$ and
requires her measure the particle $A$ with the same MB. Alice
publishes the result of the measurement, which is a sample for
eavesdropping check during the phase that the particle is
transmitted between Alice and Bob.

If Bob chooses the coding mode, he  chooses randomly one of the
four unitary operations $\{U_i\}$, say $U_B$ and performs it on
the particle $B$, and then sends the particle to Carol.

Surely, in order to prevent Alice (the server who prepares the
QIC) from eavesdropping the quantum channel between Bob and Carol,
Bob should choose the third mode, the decoy mode for the particle
$B$ with a certain probability. In this time, he replaces the
particle $B$ with the particle $d$ in state $\vert \chi\rangle_d
\in \{\vert 0\rangle, \vert 1\rangle, \vert
+x\rangle=\frac{1}{\sqrt{2}}(\vert 0\rangle + \vert 1\rangle),
\vert -x\rangle=\frac{1}{\sqrt{2}}(\vert 0\rangle - \vert
1\rangle)\}$ prepared by himself with two MBs $\sigma_z$ and
$\sigma_x$ randomly in advance, and then sends it to Carol. Bob
measures the particle $B$ with the same MB as that for preparing
the particle $d$.

(S4) Carol performs her operation on particle $B$ similar to Bob
except for the third mode. When she chooses the control mode, she
measures the particle with two MBs randomly and requires Alice do
the correlated measurement with the same MB on the particle $A$
and publish the result; otherwise she operates the particle with
one of the four unitary operations randomly, say $U_C$ and then
sends it to Alice.

(S5) Alice takes a joint Bell state measurement on the EPR pair
after she receives the particle returned from Carol. She announces
in public the result of the measurement, $U_A=U_B\otimes U_C$
which is the combined operations performed by Bob and Carol.

(S6) Carol obtains the bits encoded on the particle $B$ done by
Bob according to  her operations $U_C$ and the information
published by Alice, $R_C=U_A\otimes U_C$.

(S7) Alice, Bob and Carol repeat the processes above for
distributing the bits until they can obtain enough results $R_c$.
Bob tells Carol the position where she replaces the particle $B$
with her particle $d$.

(S8) Alice, Bob and Carol choose some of the instances as samples
for eavesdropping check and complete the analysis of the error
rates of the samples.

In detail, there are several sequences of the samples for
eavesdropping check. One is the results that Bob obtains with the
control mode, say $s_{_{Bc}}$. One is the results obtained by
Carol with the control mode, say $s_{_{Cc}}$ which are divided
into two parts, $s_{_{Cc0}}$ and $s_{_{Cc1}}$ come from the
measurements on the particle $B$ and the particle $d$
respectively. The third is the results chosen by Bob randomly from
the instances for which both Bob and Carol choose coding mode, say
$s_{_w}$. Alice, Bob and Carol exploit the refined error analysis
technique \cite{abc} for checking eavesdropping.

(S9) If all the error rates are lower than the threshold, Bob and
Carol can distill the key with error correction and privacy
amplification \cite{book} from the results $R_{BC}$ for which they
both choose the code mode. Otherwise, they abandon the results and
repeat the quantum communication from the beginning.

Now, let us discuss some issues about the security of this MUQKD
scheme.

There are two classes of eavesdroppers in this MUQKD scheme. One
is the vicious eavesdropper, Eve who does not have the access to
the particle $A$ in each EPR pair.  The other is the server, Alice
who provides the QIC for the communication and keeps the particle
$A$ in the whole process of the quantum communication. For the
former, the quantum communication between two parties in the
subsystem of this MUQKD scheme equals to Bennett-Brassard-Mermin
(BBM) QKD protocol \cite{BBM92} with or without the help of the
third parties, i.e., publishing his/her unitary operations or
results. For example, with the help of Bob's, Alice and Carol can
complete the analysis of the security of the quantum communication
in the phase that the QIC runs from Bob to Carol. The security is
embodied to the fact that the action of Eve's will disturb the
quantum systems and will be detected by Alice and Carol by
analyzing the error rate of the results $s_{_{Cc0}}$ obtained by
Carol with the control mode. From the view of eavesdropping check
for Eve, this MUQKD scheme is equal to the BBM QKD protocol
\cite{BBM92}, similar to that in the QSDC protocol \cite{twostep}.
The BBM QKD protocol is proven unconditionally secure both in
ideal condition \cite{BBMsecurity1} and in the case with noise
\cite{BBMsecurity2}. So this MUQKD scheme is secure for Eve.

As Alice has the access to the particle $A$ in each EPR pair, she
can obtain the unitary operations $U_B$ easily and will not be
detected if Bob only chooses the control mode for eavesdropping
check. That is, she performs Bell state measurement on the EPR pair
after the coding done by Bob and then gets the information without
leaving a trace. \emph{With the decoy technique}, the story will be
changed. Alice will be found out if she monitors the quantum channel
between Bob and Carol as her actions will introduce errors in the
samples $s_{_{Cc1}}$. For $s_{_{Cc1}}$ which are obtained from the
particles $d$, half of the results can be used as the samples for
eavesdropping check as the probability that Bob and Charlie choose
the same MB and then they can get the same results in principle is
50\%. If there is an eavesdropper in the line, he/she will introduce
the errors in the results as he/she does not know the MBs about the
particles $d$ and his/her action will disturb the quantum systems,
similar to that in BB84 QKD protocol \cite{BB84,book}. The error
rate introduced by an eavesdropper is 25\% if he or she monitors all
the quantum signal. The probability for choosing the decoy mode is
similar to the case with the biased bases discussed in Refs.
\cite{abc,MUQKDguo}.

There are some common features between this MUQKD scheme and that
in Ref. \cite{DLMXL}: (1) The QIC is EPR pair and the four local
unitary operations represent the different information encoded on
the states; (2) The efficiency for qubits $\eta_q$ approaches
100\% as almost all of the instance can be used as the useful
qubits and the total efficiency $\eta_t$ is 50\%; (3) Both of them
have high capacity; (4) The operations $U_C$ performed by the
receiver Carol are absolutely necessary for the QKD as they make
others know nothing about the operations $U_B$ with the results of
the combined operations published by the server Alice
$U_A=U_B\otimes U_C$ and cannot obtain the keys \cite{DLMXL}; (5)
The users need not prepare and measure the EPR pairs, and the
server provides the services; (6) The users should have the
ability for measuring a single particle with two MBs.

Of course, there are some differences in these two MUQKD schemes.
Firstly, in this scheme there is only one of the two particles in
each EPR pair running through the quantum channel, not both.
Secondly, it is unnecessary for the users to store the QIC in this
scheme, but necessary in \cite{DLMXL}. For preventing Alice from
eavesdropping the keys, Bob should exploit the decoy mode with a
certain probability. As the decoy mode is only used for
eavesdropping, the particle $d$ can be a faint laser pulse if the
QIC is photons in this scheme and any eavesdropping will be
detected \cite{BidQKD}. In this way, there is not difficulty for
Bob to prepare the particle $d$, and this scheme is easier to be
implemented than that in Ref. \cite{DLMXL}.

In  summary, we have introduced a new multi-user quantum key
distribution scheme following the ideas in quantum dense coding.
This scheme is secure if the sender of the information chooses the
decoy mode with a certain probability. It has high capacity and
its efficiency for qubit approaches 100\% as almost all the EPR
pairs can used to transmit the information. There is only one of
the two particles in each EPR pair running through the quantum
channel, and then the loss of qubits is reduced when there is loss
in the channel. Moreover, it does not require the users on the
network store the quantum states and is more convenient for
application than that in \cite{DLMXL}.

This work was supported  by the National Natural Science Foundation
of China under Grant Nos.10447106, 10435020, 10254002 and A0325401.


\begin{thebibliography}{99}
\bibitem{vernam} Vernam G S 1926  \emph{J. Amer. Inst. Elec.
Eng.} \textbf{45} 109

\bibitem{nocloning} Wootters W K and Zurek W H 1982 \emph{Nature} \textbf{
299}  802

\bibitem{BB84} Bennett C H and Brassad G 1984 \emph{Proc. IEEE Int.Conf.
on Computers, Systems and Signal Processing}  \emph{(Bangalore,
India} (New York: IEEE) PP 175-179


\bibitem{Ekert} Ekert A K 1991 \emph{Phys. Rev. Lett.} \textbf{67} 661

\bibitem{BBM92} Bennett C H, Brassard G and Mermin N D 1992 \emph{Phys. Rev.
Lett.} \textbf{68} 557

\bibitem{B92} Bennett C H 1992 \emph{Phys. Rev. Lett.} \textbf{68} 3121

\bibitem{abc} Lo H K, Chau H F  and Ardehali M 2000 \emph{Preprint} arXiv:
quant-ph/0011056

\bibitem{cabello} Cabello A 2000 \emph{Phys. Rev. Lett.} \textbf{85} 5635


\bibitem{longliu} Long G L and Liu X S 2002 \emph{Phys. Rev.} A \textbf{65} 032302

\bibitem{guoQKDE} Zhang Y S, Li C F and Guo G C 2001 \emph{Phys. Rev.} A \textbf{64} 024302

\bibitem{CORE} Deng F G and Long G L 2003 \emph{Phys. Rev.} A \textbf{68}
042315


\bibitem{guiguo} Gui Y Z, Han Z F, Mo X F and Guo G C 2003 \emph{Chin.
Phys. Lett.} \textbf{20} 608

\bibitem{hanguo} Han C, Xue P and Guo G C 2003 \emph{Chin. Phys. Lett.}
\textbf{20} 183

\bibitem{BidQKD} Deng F G and Long G L 2004 \emph{Phys. Rev.} A \textbf{70}
012311

\bibitem{delay} Deng F G, Long G L, Wang Y and Xiao L 2004
\emph{Chin. Phys. Lett.} \textbf{21} 2097
\bibitem{KZHWGTR} Kurtsiefer C, Zarda P, Halder M et. al. 2002 \emph{Nature} \textbf{419} 450

\bibitem{QKD122} Gobby C, Yuan Z L and Shields A J 2004 \emph{Appl. Phys. Lett.} \textbf{84} 3762

\bibitem{MUQKD1} Phoenix S J D, Barnett S M, Townsend P D and Blow
K J 1995 \emph{J. Mod. Opt.} \textbf{42} 1155

\bibitem{MUQKD2} Biham E, Huttner B and Mor T 1996 \emph{Phys. Rev.} A \textbf{54}
2651

\bibitem{MUQKDNature} Townsend P D 1997 \emph{Nature} \textbf{385} 47

\bibitem{MUQKDguo} Xue P, Li C F and Guo G C 2002 \emph{Phys. Rev.} A \textbf{65}
022317

\bibitem{DLMXL} Deng F G, Liu X S, Ma Y J, Xiao L and Long G L
2002 \emph{Chin. Phys. Lett.} \textbf{19} 893

\bibitem{storage1} Liu C, Dutton Z, Behroozi C H and
Hau L V 2001 \emph{Nature}  \textbf{409} 490

\bibitem{storage2} Philips D F, Fleischhauer A, Mair A et.al.  2001  \emph{Phys. Rev. Lett.} \textbf{86}
783

\bibitem{sun} Sun C P, Li Y and Liu X F 2003 \emph{Phys. Rev. Lett.} \textbf{91}
147903

\bibitem{wang} Wang K G and Zhu S Y 2002 \emph{Chin. Phys. Lett.} \textbf{19} 60

\bibitem{book} Nielsen M A and Chuang I L 2000 \emph{Quantum computation
and quantum information} (Cambridge University Press, Cambridge,
UK)



\bibitem{twostep} Deng F G, Long G L and Liu X S, 2003 \emph{Phys. Rev.} A \textbf{68}
042317

\bibitem{QOTP} Deng F G and Long G L  2004 \emph{Phys. Rev.} A \textbf{69} 052319

\bibitem{yan0} Yan F L and Zhang X Q 2004 \emph{Eur. Phys. J.} B \textbf{41} 75

\bibitem{Gaot0} Gao T, Yan F L and Wang Z X 2004 \emph{Nuovo Cimento Della Societa Italiana Di Fisica
} B \textbf{119} 313

\bibitem{GaoT1} Gao T 2004 \emph{Zeitschrift Fur Natureforshung Section} A
\textbf{59} 597

\bibitem{caiqycpl}Cai QY and Li BW 2004  \emph{Chin. Phys. Lett.} \textbf{21} 601

\bibitem{zhangzj} Zhang Z J, Man Z X and Li Y 2004 \emph{Phys. Lett.} A \textbf{333} 46

\bibitem{luh}L$\ddot{u}$ H, Yan X D, Zhang X Z 2004 \emph{Chin. Phys. Lett.} \textbf{21} 2340

\bibitem{zhangzj2} Man Z X, Zhang Z J and  Li Y 2005 \emph{Chin. Phys.
Lett.} \textbf{22} 18; 2005 \emph{Chin. Phys. Lett.} \textbf{22}
22

\bibitem{wangc} Wang C, Deng F G,  Li Y S, Liu X S and Long G L 2005 \emph{Phys. Rev.
} A \textbf{71} 044305

\bibitem{HBB99} Hillery M, Bu\v{z}ek V and Berthiaume A 1999 \emph{Phys.
Rev.} A \textbf{59} 1829

\bibitem{Peng} Li Y M, Zhang K S and Peng K C 2004 \emph{Phys. Lett.
} A \textbf{324} 420

\bibitem{densecoding} Bennett C H. and Wiesner S J 1992 \emph{Phys. Rev.
Lett.} \textbf{69} 2881


\bibitem{bf} Bostr\"{o}m K and Felbinger T  2002 \emph{Phys. Rev. Lett.} \textbf{89}
187902

\bibitem{BBMsecurity1} Inamori H, Rallan L and Vedral  V 2001 \emph{J. Phys.
} A \textbf{34} 6913

\bibitem{BBMsecurity2}Waks E, Zeevi A and Yanamoto Y 2002 \emph{Phys.
Rev.} A \textbf{65} 052310


\end{thebibliography}
\end{document}